\documentclass{article}
\usepackage{ijcai25}

\usepackage{times}
\usepackage{soul}
\usepackage{url}
\usepackage[hidelinks]{hyperref}
\usepackage[utf8]{inputenc}
\usepackage[small]{caption}
\usepackage{graphicx}
\usepackage{amsmath}
\usepackage{amsthm}
\usepackage{booktabs}
\usepackage{algorithm}
\usepackage{algorithmic}
\usepackage[switch]{lineno}

\usepackage{amssymb}
\usepackage{multirow}
\usepackage{pifont}
\usepackage[table,xcdraw]{xcolor}
\usepackage[normalem]{ulem}
\useunder{\uline}{\ul}{}
\usepackage{makecell}
\title{Can Molecular Evolution Mechanism Enhance Molecular Representation?}

\author{
Kun Li$^1$
\and
Longtao Hu$^1$\and
Xiantao Cai$^1$\and
Jia Wu$^{2}$\And
Wenbin Hu$^1$\\
\affiliations
$^1$School of Computer Science, Wuhan University, Wuhan, China\\
$^2$Department of Computing, Macquarie University, Sydney, Australia\\
\emails
\{likun98, hlt\_2003, caixiantao, hwb\}@whu.edu.cn,
Jia.wu@mq.edu.au,
}
\begin{document}

\maketitle

\begin{abstract}

Molecular evolution is the process of simulating the natural evolution of molecules in chemical space to explore potential molecular structures and properties. The relationships between similar molecules are often described through transformations such as adding, deleting, and modifying atoms and chemical bonds, reflecting specific evolutionary paths. Existing molecular representation methods mainly focus on mining data, such as atomic-level structures and chemical bonds directly from the molecules, often overlooking their evolutionary history. Consequently, we aim to explore the possibility of enhancing molecular representations by simulating the evolutionary process. We extract and analyze the changes in the evolutionary pathway and explore combining it with existing molecular representations. Therefore, this paper proposes the molecular evolutionary network (MEvoN) for molecular representations. First, we construct the MEvoN using molecules with a small number of atoms and generate evolutionary paths utilizing similarity calculations. Then, by modeling the atomic-level changes, MEvoN reveals their impact on molecular properties. Experimental results show that the MEvoN-based molecular property prediction method significantly improves the performance of traditional end-to-end algorithms on several molecular datasets. The code is available at https://anonymous.4open.science/r/MEvoN-7416/.

% Furthermore, as an emerging research paradigm, the MEvoN is valuable for molecular property prediction and can be widely applied to drug discovery areas such as molecule generation and optimization. 

% To this end, we propose a modeling approach based on the Molecular Evolution Tree (METree), which simulates atomic-level changes in molecules during their evolutionary process and explores how these changes influence molecular properties. METree constructs an evolutionary tree starting from molecules with a small number of atoms and generates evolutionary paths using similarity calculation methods. 

\end{abstract}

% Furthermore, as an emerging research paradigm, the METree is not only valuable for molecular property prediction, but also can be widely applied to drug discovery areas such as molecule generation and optimization. 
% The code is available at https://github. com/DrugD/METree.

\section{Introduction}

% The discovery of new drugs plays a pivotal role in addressing global health challenges, and the efficient prediction of molecular properties is a critical step in this process. The process of drug discovery involves identifying molecular candidates with desirable properties, such as potency, selectivity, and minimal toxicity. However, the vast chemical space and complexity of molecular interactions make this task extremely challenging. A fundamental aspect of drug discovery is the accurate representation of drug molecules, as their structure directly correlates with their properties and biological activities. Therefore, the development of effective molecular representation methods is essential for improving the efficiency of drug discovery and materials science.

\begin{figure}[ht!]
    \centering
    \includegraphics[width=0.99\linewidth]{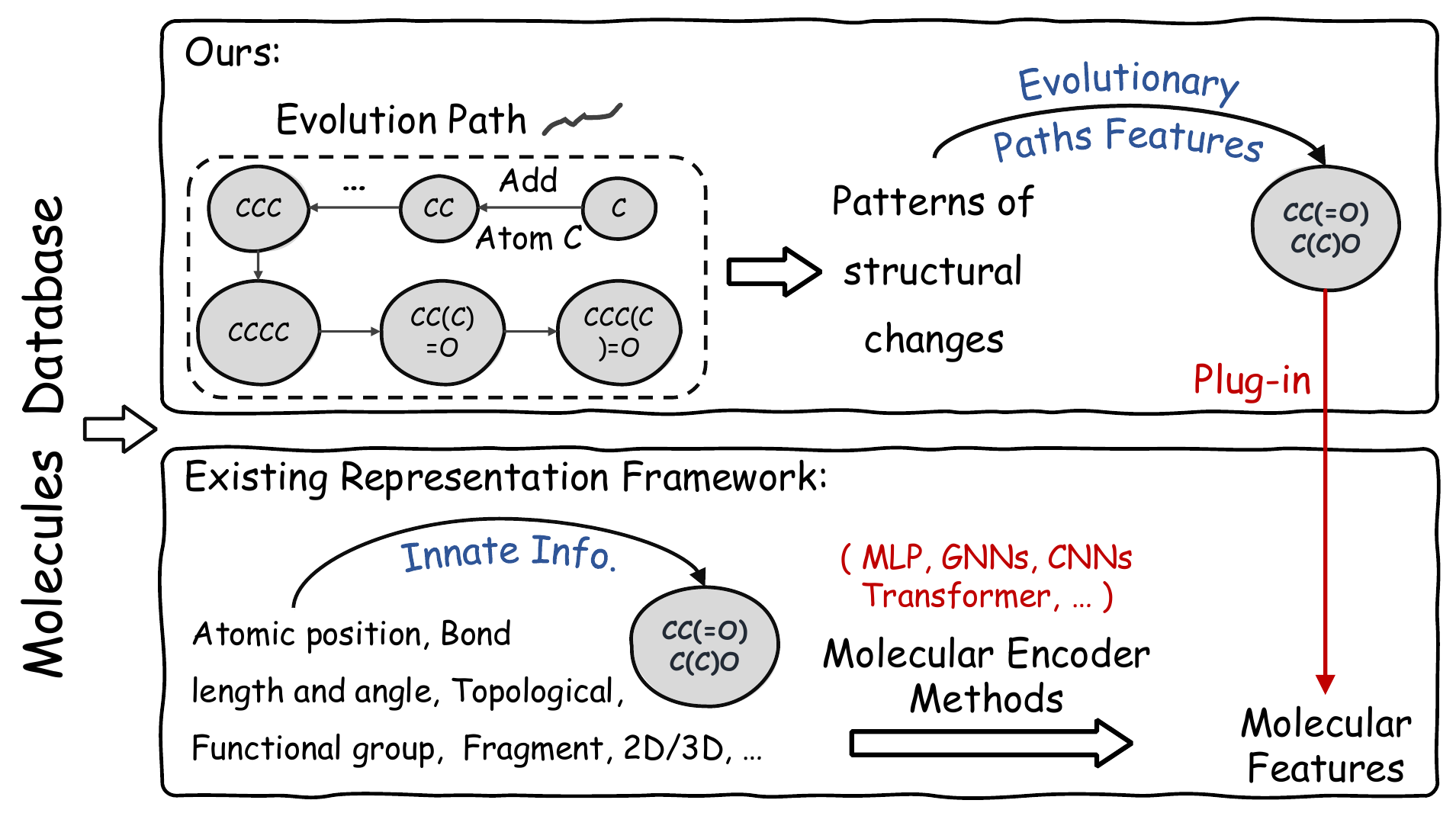}
    \caption{Molecular evolutionary network (MEvoN) illustrating the evolution pathway, providing a quantitative method for assessing the magnitude and direction of changes in molecular properties.}
    \label{fig:mov}
\end{figure}

Molecular evolution is the process of exploring potential  molecular structures and properties by simulating the evolution of molecules in nature using structural mutations (e.g., substitutions, additions, deletions and isomerization) to make the molecules evolve in the chemical space \cite{CST,lameijer2006molecule}. This concept is widely applied in molecular generation and optimization \cite{adelusi2022molecular} for novel chemical structure discovery and potential active molecule identification. For example, by simulating Darwinian evolution through crossover and mutation, genetic algorithms \cite{fu2022reinforced} continuously optimize molecular structures to find optimal solutions in vast chemical space. Chemical space travel,  proposed by Ruud van Deursen and Jean-Louis Reymond \cite{CST}, combines molecular evolution with algorithms to efficiently obtain target molecules through multiple mutations that start with an initial molecule. For larger molecules, protein changes are interconnected within a multi-dimensional space through mechanisms such as mutations. In this scenario, random mutation and natural selection cooperate to shape the structure and function of larger molecules \cite{memtgr}. For instance, the ESM3 multimodal language model can simulate this natural evolutionary process \cite{ESM3}, significantly enhancing the model's analytical and inference capabilities. Hence, this evolutionary mechanism improves our understanding of the relationship between molecular structure and biological activity and leverages the structural variation patterns among similar molecules.

\begin{figure}[t!]
    \centering
    \includegraphics[width=1\linewidth]{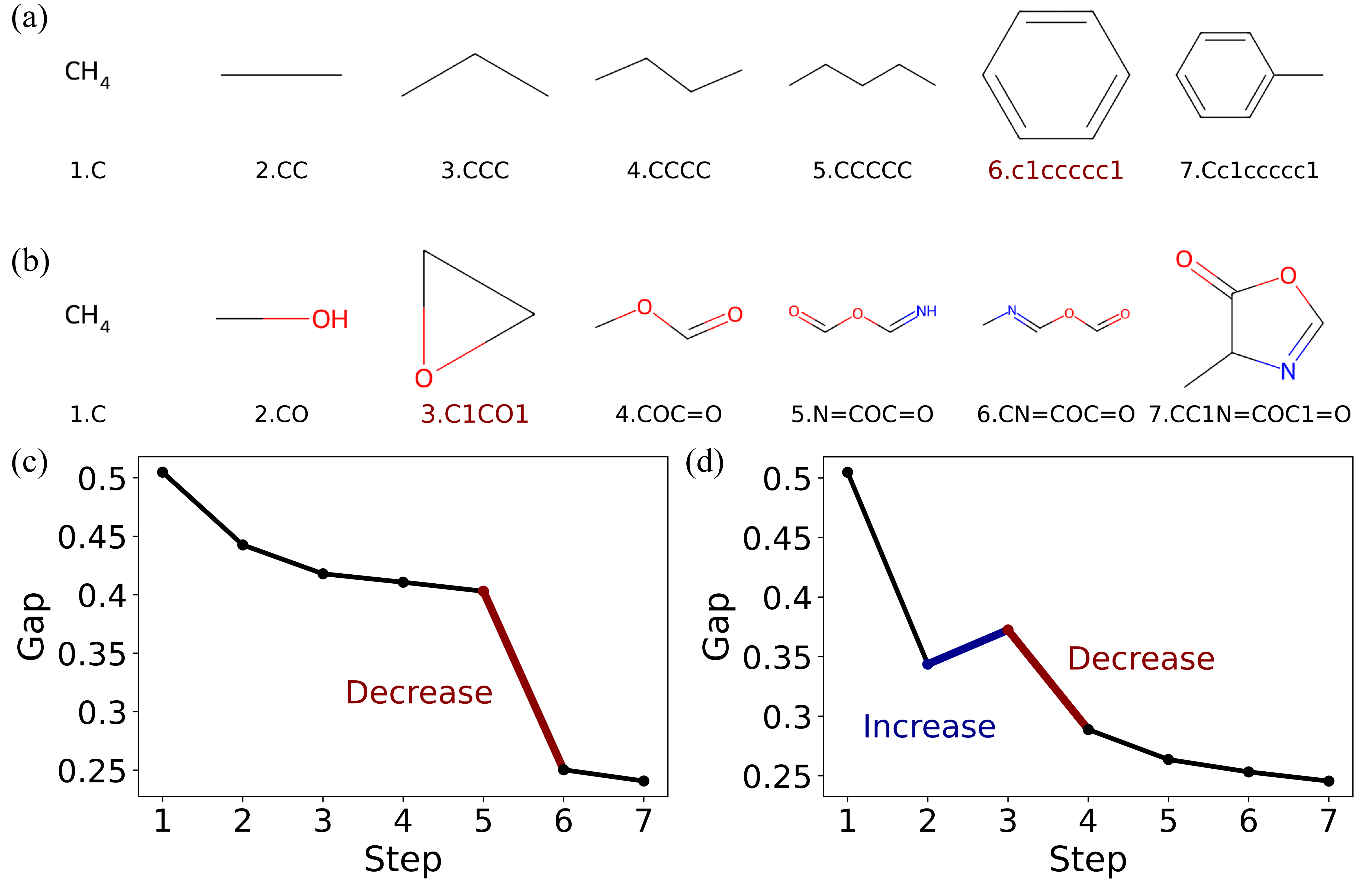}
    \caption{Evolutionary paths and molecular property changes for two molecules from the QM9 dataset. (a) and (c) correspond to 'Cc1ccccc1', while (b) and (d) correspond to 'CC1N=COC1=O'. (a) and (b) illustrate the evolutionary paths of two molecules. (c) and (d) display the corresponding variations in molecular properties.}
    \label{fig:molecule_path_case}
\end{figure}

% However, existing molecular representation methods primarily focus on extracting atomic-level structures and chemical bonds from molecules themselves \cite{wang2024enhancing,satorras2021n,schutt2017schnet},

Molecular representation methods based on graphs, sequences, and fingerprints have been widely used in drug screening \cite{moshkov2023predicting,drugdiscovery1}, materials science \cite{born2023regression,trabucco2022design}, and molecular design \cite{Prompt-MolOpt,regressor-free,li2024fragment}. During drug discovery, graph- and sequence-based methods \cite{sharmadiffuse,ijcai2024p234,ijcai2024p235,tcnns,11dGCN} are used to screen potential candidates from large molecular datasets. As shown in Figure \ref{fig:mov}, existing molecular representation methods extract innate information from molecules \cite{wang2024enhancing,satorras2021n,schutt2017schnet}, such as atomic structures, chemical bonds, or other two-and three-dimensional features, often overlooking their historical evolutionary paths. This raises the question: can we enhance molecular representations by simulating the molecule evolution process, and extracting and analyzing the structural changes within the evolutionary pathway? By extracting and analyzing structural changes along evolution pathways, we may gain deeper insights into their influence on molecular representations, and integrating these findings with existing methods can enhance the overall understanding and depiction of molecular properties.

Notably, similar molecules inherently contain rich information, and their property variations often follow certain trends. Therefore, we conduct an evolutionary analysis on molecules from the QM9 dataset, using the molecular orbital gap (i.e., HOMO-LUMO gap) as the target property. We focus on molecules with high similarity and a single atom difference. Figure \ref{fig:molecule_path_case} illustrates two molecular evolutionary paths: "Cc1ccccc1" and "CC1N=COC1=O," highlighting these trends. The property variation curves in Figure \ref{fig:molecule_path_case} (a) exhibit a sharp decrease between steps 5 and 6, when a carbon chain transforms into a benzene ring. This is because an increase in molecular size normally leads to a decrease in the energy gap. In addition, closing the carbon chain into a benzene ring introduces additional electron delocalization through its aromatic structure, significantly reducing the HOMO-LUMO energy gap \cite{cornil2001interchain}. This aligns with the general trend that aromatic molecules tend to have lower energy gaps \cite{pope1999electronic}. Similarly, as shown in Figure \ref{fig:molecule_path_case} (d), significant changes in the evolutionary path of 'CC1N=COC1=O' occur between steps 2–3 and 3–4. These steps involve the formation and breaking of a 3-membered ring due to its high energy strain. Cleaving a 3-membered ring releases strain energy and alters the electronic structure, significantly impacting energy levels \cite{planells2020accurate}. By analyzing these mutation effects and patterns, we gain a deeper understanding of the relationship between molecular structure and properties.

\begin{figure*}[t!]
    \centering
    \includegraphics[width=0.85\linewidth]{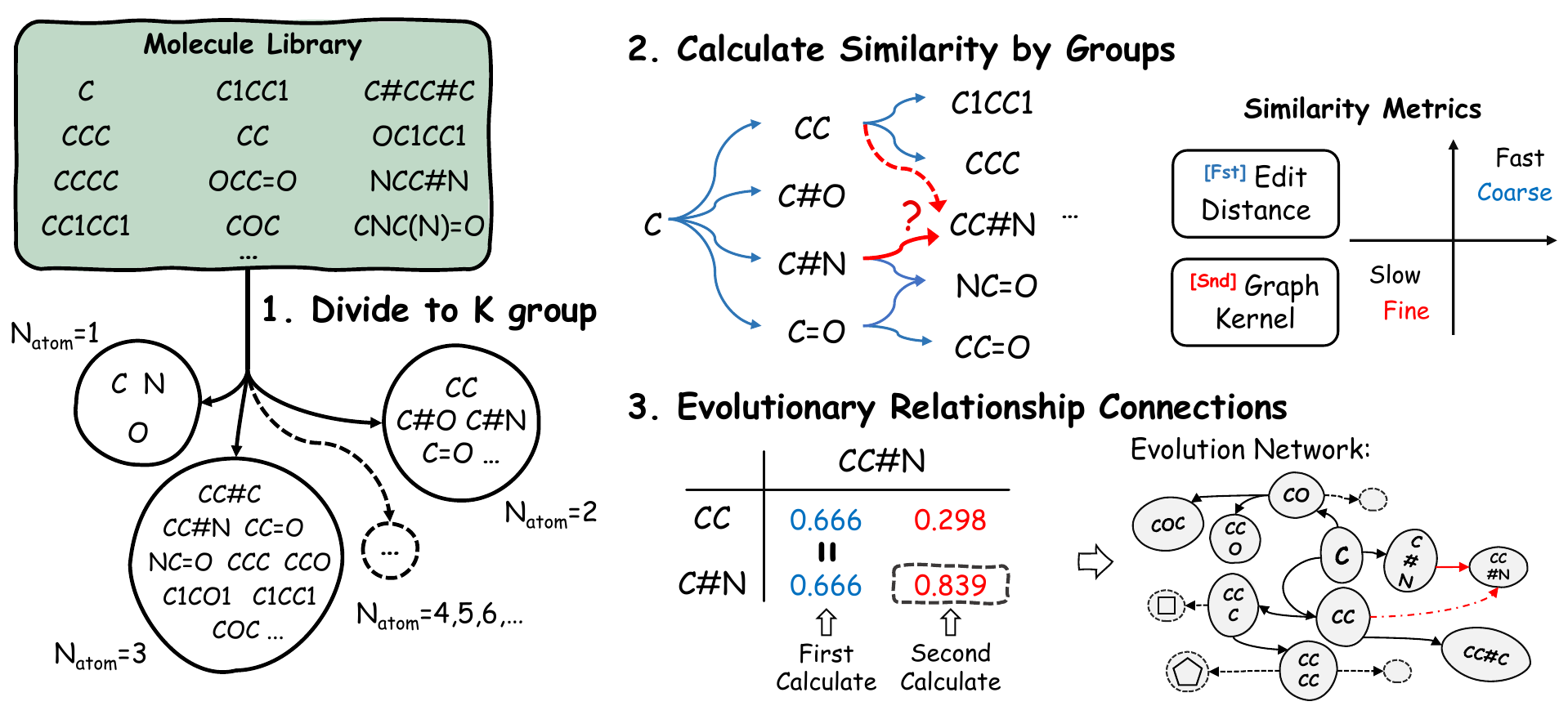}
    \caption{The MEvoN method's construction process. The steps are: 1) group molecules by atom count; 2) calculate inter-group similarity; and 3) determine evolutionary relationships based on multiple similarity measures.}
    \label{fig:overview}
\end{figure*}

Therefore, we explore the possibility of employing the phylogenetic analysis methods used in genomics and protein sequencing to construct a network that describes molecule evolution. This network can simulate atomic-level changes during molecular evolution. As a result, we propose the \underline{\textbf{m}}olecular \underline{\textbf{evo}}lutionary  \underline{\textbf{n}}etwork (MEvoN) for molecular representations. MEvoN regards molecules with fewer atoms as ancestral nodes and those with more as descendants. Thus, we can construct the evolutionary relationships by calculating the similarity between these two molecular node types. The MEvoN is formed by various evolutionary paths and molecular node sets, revealing the impact of atomic-level changes on molecular properties. Furthermore, we demonstrate the application of the MEvoN-based molecular property prediction method (MEvoN-MPP). The MEvoN-MPP method combines the evolutionary path- and label-aware modules to effectively capture the evolutionary information. The experimental results demonstrate that MEvoN-MPP, as a basic property prediction  model plug-in, effectively integrates the molecule's evolutionary path information with the inherent features to enhance its representation. This paper's contributions are as follows:

\begin{itemize}
    \item A novel molecular representation paradigm based on the evolutionary network is proposed. Evolutionary relationships are constructed by calculating the similarities between molecules, thus helping to analyze the influence of atomic-level changes on molecular properties.
    \item To integrate evolutionary information with molecular features, we propose the MEvoN-MPP method. Experiments on the several datasets indicate that our method improves molecular representation by an average of 32.3\%, validating MEvoN's effectiveness.
    % \item We propose the MEvoN-MPP for molecular property prediction, which combines evolutionary information with molecular structural features to enhance molecular representation capabilities.
    % % \item We present the MEvoN-MPP for molecular property prediction, which combines evolutionary information with molecular structure features, improving property prediction models'  performance.
    % \item Experiments on MEvoN construction and molecular property prediction were conducted across multiple datasets. The results demonstrate the effectiveness of MEvoN in molecular property prediction and highlight its broad potential applications in drug discovery.
\end{itemize}

% METree construction and METree-MPP experimental tests are performed on multiple datasets. The results show that METree-MPP outperforms traditional end-to-end algorithms, verifies the effectiveness of the METree in molecular property prediction, and demonstrates the potential of METree to be widely used in drug discovery.

% \section{Preliminaries}

% We use $\odot$ to denote the element-wise (Hadamard) product, $\| \cdot \|_2$ to denote the row-wise L2 norm, and $\times$ to represent the cross-product between vectors. The root of the tree is usually defined as the most basic or ancestral molecule, which can be a simple atom or molecule (e.g., carbon). The tree is constructed iteratively by selecting the most relevant evolutionary steps that connect molecules with similar properties.

% Finally, the tree can be visualized with molecular diagrams at each node to highlight the evolution of molecular structures and their corresponding properties.

\section{Molecule Evolutionary Network}
\label{sec:MEvoN}
% In this section, we describe the architecture of the model and the various components used for processing graphs and sequences. Our model is designed to handle drug molecule graph data, path information, and sequence data using a combination of Graph Neural Networks (GNN), Transformers, and fully connected layers.

In this section, we systematically describe the method and principles for constructing the MEvoN. Figure \ref{fig:overview} presents the MEvoN model's construction process.  First, the molecules are grouped according their atom count. Then, the similarity calculations and evolutionary relationship constructions are performed between molecules from different groups, as described in Algorithm \ref{alg:MEvoN}. 

% Molecules are grouped based on their atom count, with similarity calculations and the construction of evolutionary relationships performed between molecules from different groups, as described in Algorithm \ref{alg:MEvoN}.

% This method allows us to determine the evolutionary relationships between molecules, ultimately constructing the MEvoN.

\paragraph{Notations.}
We formulate the MEvoN as a network representation $\mathcal{N}  = (\mathcal{V}, \mathcal{E} )$. The molecules are represented as the set of node $\mathcal{V}$, and $\mathcal{E}$ is the set of edges that connects the nodes. The set of nodes $\mathcal{V}$ contains $\mathrm{N}$ molecules, and the corresponding properties of each molecule are represented by $P_i \in \mathcal{P}$. Furthermore, each  MEvoN is constructed from the molecular dataset with $\mathcal{M}$ representing the set of all the molecules from one dataset.

% The degree of a node, denoted by $k$, represents the number of edges connected to a node, which corresponds to the number of evolutionary branches diverging from that molecule. 

% \subsection{Overview}
% 分子进化树（METree）的构建旨在探索分子之间的进化关系，追溯其在结构和成分上的变化。构建过程从一个初始分子开始，例如碳（C），根据分子中原子的数量和种类的不同，树的层级会有所不同。每个层级代表了分子进化的不同阶段，反映了分子结构在演化过程中逐步变化的过程。通过这种层级结构，能够清晰地展示分子从最基础的形态到更复杂结构的演化路径。

% 进化树的结构是基于进化距离或遗传变异度来构建的，进化距离量化了分子之间的差异程度。进化距离越短，表示分子之间的相似度越高，经历的变化越少或变化较小；而进化距离越长，则表示分子之间的差异越大，经历了更多或更显著的变化。在构建过程中，分支代表分子转化的过程，例如原子的添加、去除或替代，而节点表示进化过程中不同阶段的分子。这种结构使得分子之间的亲疏关系得以可视化，揭示了分子在进化过程中如何发生变化，并帮助理解不同分子之间的演化联系。

\subsection{Molecular Grouping}

The MEvoN's construction aims to explore the evolutionary relationships among molecules by tracing their structural and compositional changes. To facilitate the MEvoN's construction, molecules are grouped according to their atomic compositions. Thus, the number of atoms in each molecule serves as a distinguishing feature, organizing them  into hierarchical groups. Molecules with fewer atoms form the base or initial network stages, while those with more  appear  later.

During the grouping process, the number of atoms $\mathrm{N}_{\text{atoms}}$ in each molecule is a positive integer. This constraint ensures that each molecule can be uniquely categorized according to its atomic count. The grouping process involves molecular iterations within the dataset (typically represented by SMILES strings). It also involves extracting the number of atoms $\mathrm{N}_{\text{atoms}}(\cdot)$ in each molecule, and categorizing them into the corresponding atom count groups $G_k$. These groups serve as the evolutionary network's initial levels. Formally, the molecules are grouped as:
\begin{equation}
G_k = \{ m_i \mid \mathrm{N}_{\text{atoms}}(m_i) = k, \, m_i \in \mathcal{M} \},
\end{equation}
\noindent where $m_i$ denotes a molecule and $\mathrm{N}_{\text{atoms}}(m_i)$ is the number of atoms in $m_i$. 

The grouping process begins with the initial molecules, such as the molecule with one C atom. Molecules with the same number of atoms $k$ are grouped into the set $G_{k}$, representing different stages of molecular evolution. The purpose of the molecular grouping is to simulate the gradual increase in atom count that is typically observed in natural molecule evolution. Therefore, evolutionary relationships are not constructed between molecules within the same group $G_{k}$.

\subsection{Inter-Group Similarity Calculation}

The MEvoN method construction's core lies in calculating evolutionary distance or molecular similarity to establish the evolution pathways. The similarity between two molecules, $ m_i $ and $ m_j $, denoted as $ S(m_i, m_j)$, can be measured using various similarity metrics. The similarity value is between $[0,1]$, where 1 indicates complete similarity and 0 implies no similarity.

% Commonly used fingerprint types include Morgan fingerprints, MACCS keys, AtomPair fingerprints, and TopologicalTorsion fingerprints.
 
\begin{itemize}
    \item \textit{Fingerprint-based similarity}: Fingerprint-based similarity methods \cite{wang2024predicting} represent molecules as binary fingerprint vectors, where each bit indicates the presence or absence of a specific structural feature within the molecule. The Tanimoto coefficient~\cite{tanimoto} is the most widely used similarity measure, quantifying the overlap between two binary fingerprints as follows:
    \begin{equation}
    S_{\text{fp}}(m_i, m_j) = \frac{|F(m_i) \cap F(m_j)|}{|F(m_i) \cup F(m_j)|},
    \end{equation}
    where $F(m_i)$ and $F(m_j)$ represent the fingerprint sets of molecules $m_i$ and $m_j$, respectively. 
    % The Tanimoto similarity $\mathcal{S}_{\text{FP}}$ ranges from 0 to 1, where 1 indicates complete similarity and 0 indicates no overlap.
    \item \textit{Graph-based similarity}: Graph-based similarity is determined by comparing the graphs using graph kernels, such as the Weisfeiler–Lehman graph kernel $\mathcal{S}_{\text{wl}}(m_i, m_j)$~\cite{wl}, described as follows:
    \begin{equation}
    \mathcal{S}_{\text{wl}}(m_i, m_j) = \text{WL}(G(m_i), G(m_j)),
    \end{equation}
    where $G(m_i)$ and $G(m_j)$ represent the graph representations of molecules $m_i$ and $m_j$, respectively.

    \item \textit{Edit distance similarity}: The molecular edit distance $d_{\text{edit}}(m_i, m_j)$ measures the number of changes required to convert one molecule into another, where the changes correspond to atom insertions, deletions, and substitutions. The edit distance is given by:
    \begin{equation}
    \mathcal{D} _{\text{edit}}(m_i, m_j) = \min_{\Theta} \left( \sum_{opt \in \Theta} \text{cost}(opt) \right),
    \end{equation}
    where, $\Theta$ represents the set of possible edit operations (i.e., insertions, deletions, and substitutions), and each $opt$ is the cost associated with a specific operation. Thus, the edit distance similarity \( \mathcal{S}_{\text{edit}}(m_1, m_2) \) is defined as:
    \begin{equation}
    \mathcal{S}_{\text{edit}}(m_1, m_2) = 1 - \frac{\mathcal{D}_{\text{edit}}(m_1, m_2)}{\max(\text{len}(m_1), \text{len}(m_2))},
    \end{equation}
\end{itemize}

The similarity measure $S(m_i, m_j)$ ranges from 0 to 1. A higher value indicates greater similarity, capturing molecular structural differences and feature changes. To ensure consistency and clarity along the evolutionary pathway, the similarity calculation is only conducted when the inter-group condition $i < j$ met. Specifically, similarity is calculated only between $m_i \in G_i$ and $m_j \in G_j$. As a result, the inter-group similarity calculation effectively avoids redundant computations, ensuring that the evolutionary path progresses effectively within the hierarchical structure.

% During the construction of the evolutionary tree, the branches represent molecular transformations, such as the addition, deletion, or substitution of atoms, while the nodes represent molecules at different evolutionary stages.

\begin{algorithm}[t!]
    \caption{MEvoN Construction Algorithm}
    \label{alg:MEvoN}
    \textbf{Input}: Set of molecules $\mathcal{M}$. \\
    \textbf{Parameter}: Similarity thresholds $\theta_1$ and $\theta_2$. \\
    \textbf{Output}: A MEvoN $\mathcal{N}= (\mathcal{V},\mathcal{E})$.
    
    \begin{algorithmic}[1]
        \STATE \textbf{Group} molecules by atomic count: 
        \STATE \hspace{2em} $G_k = \{ m_i \mid \mathrm{N}_{\text{atoms}}(m_i) = k, m_i \in \mathcal{M} \}$
        
        \FOR{each pair of evolutionary groups $(G_i, G_j)$, $i < j$}
            \STATE \textbf{Initialize:} $\text{Pair}^1 = \emptyset$, $\text{Pair}^2 = \emptyset$
            \FOR{each molecule pair $(m_p \in G_i, m_q \in G_j)$}
                \STATE Calculate the similarity $S_{\text{edit}/\text{fp}}(m_p, m_q)$
                \IF{$S_{\text{edit}/\text{fp}}(m_p, m_q) \geq \theta_1$}
                    \STATE Add $(m_p, m_q)$ to $\text{Pair}^1$
                \ENDIF
            \ENDFOR
            
            \STATE \textbf{Let} $\text{Pair}^1_{\text{max}}$ be the maximum similarity pairs in $\text{Pair}^1$
            \IF{$\left \| \text{Pair}^1_{\text{max}} \right \|  > 1$}
                \FOR{each pair $(m_p^\prime, m_q^\prime) \in \text{Pair}^1_{\text{max}}$}
                    \STATE Calculate the similarity $\mathcal{S}_{\text{wl}}(m_p^\prime, m_q^\prime)$
                    \IF{$\mathcal{S}_{\text{wl}}(m_p^\prime, m_q^\prime) \geq \theta_2$}
                        \STATE Add $(m_p^\prime, m_q^\prime)$ to $\text{Pair}^2$
                    \ENDIF
                \ENDFOR
            \ENDIF
            
            % \STATE \textbf{Update METree:}
            % \STATE Add new nodes $\mathcal{V}^\prime$ and edges $\mathcal{E}^\prime$ from $\text{Pair}^2$ to the tree:
            \STATE $\mathcal{V}^\prime \gets \left\{ m_p^\prime, m_q^\prime \mid (m_p^\prime, m_q^\prime) \in \text{Pair}^2 \right\}$
            \STATE $\mathcal{E}^\prime \gets \left\{ (m_p^\prime, m_q^\prime) \mid (m_p^\prime, m_q^\prime) \in \text{Pair}^2 \right\}$
            \STATE $\mathcal{N} \gets (\mathcal{V} \cup \mathcal{V}^\prime, \mathcal{E} \cup \mathcal{E}^\prime)$
        \ENDFOR
    \end{algorithmic}
\end{algorithm}

\subsection{Establishing Evolutionary Relationships}

After calculating the molecular similarities, valid edges $ \mathcal{E}$ are added to the MEvoN to represent evolutionary relationships. These edges are formed based on the similarity values $S(m_i, m_j)$ between molecule pairs. Then, the predefined thresholds $\theta_1$ and $\theta_2$ are used to determine whether two molecules are evolutionarily related. 

% Specifically, an edge $ \mathcal{E}_{i,j}$ is created if $S(m_i, m_j) \geq \theta$.

% if the similarity score between two molecules exceeds the thresholds, an edge is created, indicating that the two molecules share a significant evolutionary relationship. Formally,
 
Consider two evolutionary groups, $G_i$ and $G_j$, where $i < j$. This implies that $G_i$ represents molecules from an earlier evolutionary stage, and $G_j$ denotes those from a later phase. To calculate the similarity between $G_i$ and $G_j$, each molecule $m_q \in G_j$ is compared with every $m_p \in G_i$, and the similarity between $m_q$ and $m_p$ is computed.

To establish evolutionary relationships, the $\mathcal{S}_{\text{edit}/\text{fp}}$ similarity function is used to calculate molecular similarity:
\begin{equation}
\text{Pair}^1 = \left\{ (m_p, m_q) \mid \mathcal{S}_{\text{edit}/\text{fp}}(m_p, m_q) \geq \theta_1 \right\},
\end{equation}

Thus, we obtain $\text{Pair}^1$ as the result of the first-stage screening. Let the maximum value in $\text{Pair}^1$ be denoted as $\text{Pair}^1_{\text{max}} = \left\{ \left( m_p, m_q \right) \, | \, \mathcal{S}_{\text{edit}/\text{fp}}(m_p, m_q) = \mathrm{Max} \left( \text{Pair}^1 \right) \right\}$.  In this scenario, $\mathcal{S}_{\text{edit}/\text{fp}}$  is the quickest and most efficient measure. However, a second round of similarity calculation is required to obtain a more precise evolutionary relationship, when the number of elements in $\text{Pair}^1_{\text{max}}$ is greater than one.

% In the second round, a more precise similarity calculation is performed using the $\mathcal{S}_{\text{wl}}$. The graph-based measure captures more complex topological similarities between molecules, which makes it more accurate in distinguishing subtle differences in molecular structure. 
% In the second round of calculations, only those molecule pairs whose graph-based similarity exceeds a higher threshold $\theta_2$ are considered to have sufficient similarity to establish an evolutionary relationship.

In the second round, a more precise similarity calculation is conducted using the $\mathcal{S}_{\text{wl}}$ operation. This graph-based measure captures more intricate topological similarities between molecules, enhancing its ability to distinguish subtle structural differences, expressed as:
\begin{equation}
\text{Pair}^2 = \left\{ (m_p^\prime, m_q^\prime) \mid \mathcal{S}_{\text{wl}}(m_p^\prime, m_q^\prime) \geq \theta_2, \right\},
\end{equation}
\noindent where $(m_p^\prime, m_q^\prime) \in \text{Pair}^1_{\text{max}}$ and $\text{Pair}^2$ is the final result. With the molecule pair $\text{Pair}^2$, a new set of nodes $\mathcal{V}^\prime$ and edges $\mathcal{E}^\prime$ can be added to $\mathcal{N}$:
\begin{align}
\left\{
\begin{aligned}
\mathcal{V}^\prime &= \left\{ m_p^\prime, m_q^\prime \mid (m_p^\prime, m_q^\prime) \in \text{Pair}^2 \right\}, \\
\mathcal{E}^\prime &= \left\{ (m_p^\prime, m_q^\prime) \mid (m_p^\prime, m_q^\prime) \in \text{Pair}^2 \right\},
\end{aligned}
\right.
\end{align}

Thus, the evolutionary network $\mathcal{N}$ is updated by incorporating the new nodes and edges:
\begin{equation}
\mathcal{N} = (\mathcal{V} \cup \mathcal{V}^\prime, \mathcal{E} \cup \mathcal{E}^\prime).
\end{equation}
\noindent where, $\mathcal{V}$ and $\mathcal{E}$ represent the original set of nodes and edges, while $\mathcal{V}^\prime$ and $\mathcal{E}^\prime$ represent the newly added ones.

\begin{figure}[t!]
    \centering
    \includegraphics[width=0.9\linewidth]{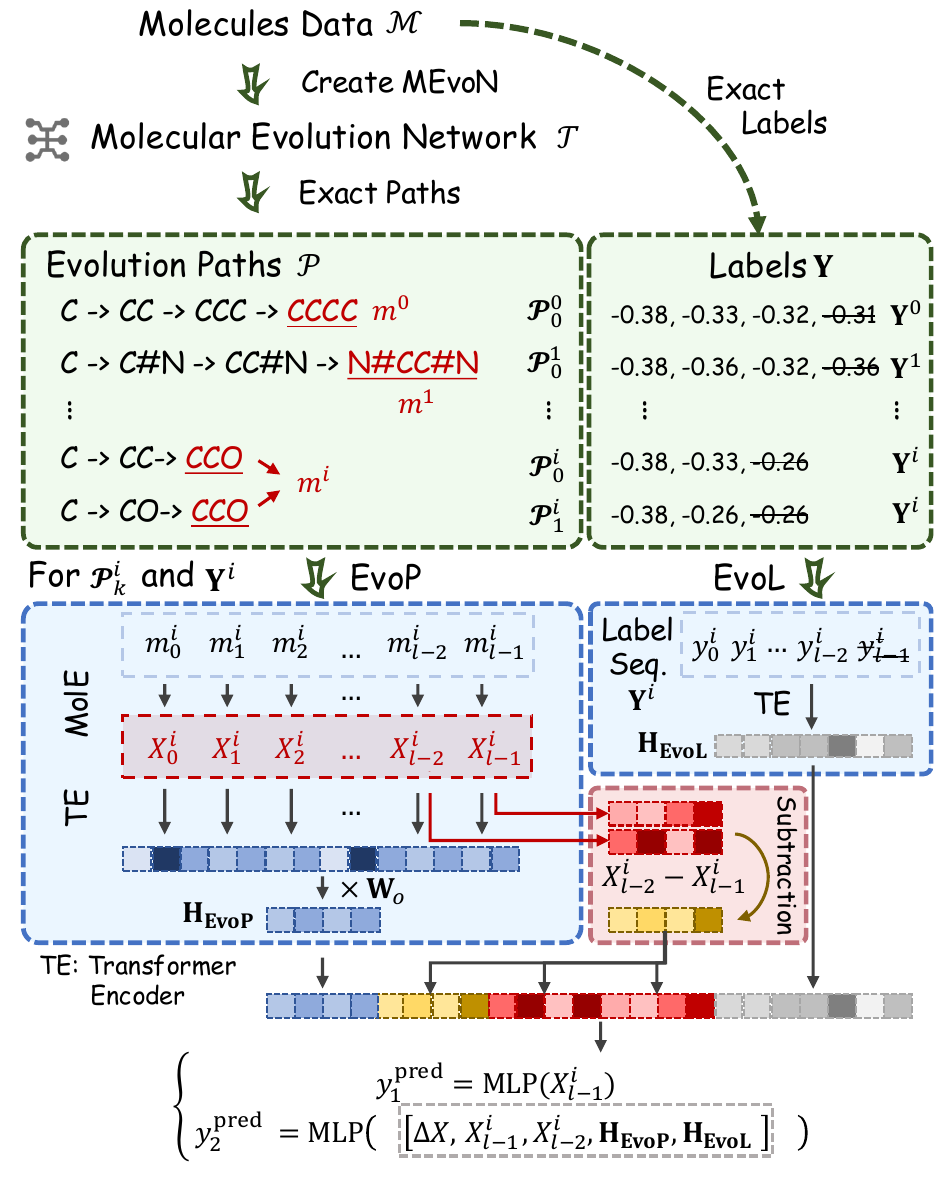}
    \caption{Overview of MEvoN-MPP, which employs the MEvoN method to predict various molecular properties. The process includes evolutionary feature extraction and property prediction.}
    \label{fig:MEvoNMPP}
\end{figure}

\section{Molecular Property Prediction Using MEvoN}

% In the process of drug discovery and molecular design, accurately predicting the physicochemical properties and biological activities of molecules is crucial. Traditional molecular property prediction methods often rely on the structural features of molecules and empirical data, but they tend to overlook the historical evolution of molecules.

The evolutionary relationships between molecules provide valuable contextual information for understanding structure-property dependencies. By leveraging MEvoN, we can incorporate the molecules' evolutionary paths as auxiliary information, thereby enhancing representation. Therefore, we propose the MEvoN-MPP model, a MEvoN-based molecular property prediction method. MEvoN-MPP includes the path- (EvoP) and label-aware (EvoA) modules, and the molecular encoder (MolE). The EvoP module captures the evolutionary relationships between molecules, while the EvoA module leverages label information to weight the evolutionary paths, enabling the model to understand each molecule's evolutionary context more effectively. MolE encodes the structural features from molecular graphs and can utilize any deep learning model capable of encoding molecules, such as Graph Neural Networks (GNNs), Convolutional Neural Networks (CNNs), and Transformers. The steps of MEvoN-MPP are as follows:

First, let $\mathcal{N}$ be the MEvoN constructed from a set of molecules $\mathcal{M}$ (see Section \ref{sec:MEvoN}). For a single molecule $m_i$, we locate its position within the MEvoN and trace its evolutionary paths (denoted as $\mathcal{P}$). The evolutionary path $\mathcal{P}^i$ refers to the set of paths from the network's root node to the molecule $m_i \in \mathcal{M}$, which can be collected by the backtracking algorithm.  Specifically, $\mathcal{P}^i$ is a path set, where each path $\mathcal{P}^i_k= \left ( m_0^i, m_1^i, \dots, m_{l-1}^i, m_l^i \right )$ represents an evolutionary path from the root to $m_i$. The length of path $\mathcal{P}^i_{k}$ denotes $l$, i.e., the number of molecules included in the path. Each molecule's graph features are extracted with MolE and serve as the initial path features for $\mathcal{P}^i_k$. This can be expressed as:
\begin{equation}
\mathbf{P}^i_k = \left [ \mathrm{MolE}(m_0^i), \dots , \mathrm{MolE}(m_l^i) \right ],
\end{equation}
\noindent where $\left [ \cdot \right ] $  denotes element concatenation. Then, we obtain the path features $\mathbf{P}^i \in \mathbb{R}^{K \times L \times F}$, where $K$ is the number of evolutionary paths $\mathcal{P}^i$, $L$ is the maximum path length, and $F$ is the feature dimension of each molecule obtained from the MolE. Thus, we encode the $\mathbf{P}^i$ as follows:
\begin{align}
\left\{
\begin{aligned}
\mathbf{H}_{\text{pos}} &= \mathbf{W}_e \mathbf{P}_i + \mathbf{b}_e + \mathbf{E}_p, \\
\mathbf{H}_{\text{out}} &= \text{TransformerEncoder}(\mathbf{H}_{\text{pos}}),
\end{aligned}
\right.
\end{align}
\noindent where $\mathbf{W}_e \in \mathbb{R}^{F \times D}$ is the weight matrix, $\mathbf{b}_e \in \mathbb{R}^D$ denotes the bias vector, and $D$ represents the embedding dimension. To incorporate sequential dependencies, we incorporate learnable positional encoding $\mathbf{E}_p \in \mathbb{R}^{L \times D}$ into the embeddings. After that, the position embeddings $\mathbf{H}_{\text{pos}} \in \mathbb{R}^{K \times L \times  D}$ are then passed through a Transformer encoder to capture the dependencies among molecules, producing the output sequence $\mathbf{H}_{\text{out}} \in \mathbb{R}^{K \times L \times  D}$.

% \begin{table}[t!]
%     \centering
%     \renewcommand\arraystretch{1.1} % 行间距
%     \setlength{\tabcolsep}{5mm}{
%     \resizebox{\columnwidth}{!}{%
%     \begin{tabular}{ccccc}
%     \hline
%     Dataset & Molecules & Edges & Nodes & Max Path \\ \hline
%     QM7 & 6832 & 9095 & 6832 & 110 \\ \hline
%     QM8 & 21766 & 27068 & 21766 & 46 \\ \hline
%     QM9 & 133885 & 165790 & 133330 & 253 \\ \hline
%     \end{tabular}}}
%     \caption{Overview of MEvoN construction on the QM7, QM8, and QM9 datasets.}
%     \label{tab:qm_datasets_summary}
% \end{table}

Subsequently, the final prediction is obtained by selecting the last hidden state $\mathbf{h}_{\text{last}} \in \mathbb{R}^{K \times 1 \times D}$ from the output sequence, which  is passed through a fully connected layer to produce the predicted molecular property $\mathbf{H}_{\text{EvoP}}$:
\begin{equation}
\mathbf{H}_{\text{EvoP}} = \mathbf{W}_o \mathbf{h}_{\text{last}} + \mathbf{b}_o,
\end{equation}
\noindent where $\mathbf{W}_o \in \mathbb{R}^{D \times 1}$ is the weight matrix and $\mathbf{b}_o  \in \mathbb{R}^{K}$ is the bias term.

The EvoL module's computation is similar to that of EvoP. The EvoL input is the molecular properties in $\mathcal{P}^i_k$, denoted as $\mathbf{Y}^i\in\mathbb{ R }^{L}$. In this case, $y_{l -1 }^i $ and $ y_{l -2 }^i $ represent the labels of the last and  the second-to-last valid molecules of $\mathbf{Y}^i$, respectively. The label of each path's last valid molecule, $y_{l-1 }^i$, is masked to prevent data leakage. After encoding with EvoL, the label path feature $\mathbf{H}_{\text{EvoL}} $ is obtained, which can be expressed as $\mathbf{H}_{\text{EvoL}} = \text{EvoL}(\mathbf{Y}^i)$.

For path $ \mathcal{P}^i_k $, the last two valid molecules, $ m_{l-2}^i$ and $m_{l-1}^i $, serve as input to the MolE. MEvoN-MPP predicts the property changes caused by the molecular pair, learning their evolution patterns—specifically, the property changes arising from the addition of atoms and chemical bonds at different positions. Then, the Evo and the Mol branches are used to predict the property changes caused by the molecular pair $ (m_{l-2}^i$, $m_{l-1}^i) $ and the properties of $m_{l-1}^i$, respectively. 

In the MolE branch, property prediction is performed directly on the feature  $X_1$ extracted by the MolE, and the output is denoted as $y_1^{\text{pred}}$. The molecular representations $X_1$ and $X_2$ can be expressed as:
\begin{equation}
X_1  = \mathrm{MolE}(m_{l-1}^i), \;\;\;X_2 = \mathrm{MolE}(m_{l-2}^i),
\end{equation}

% For each path, the last valid molecule, $m_i$, undergoes masking to prevent data leakage. After encoding with EvoL, the label path feature $\mathbf{H}_{\text{EvoL}}$ is obtained, which can be expressed as $\mathbf{H}_{\text{EvoL}} = \text{EvoL}(\mathbf{Y}^i)$.

% For each path $\mathcal{P}^i_j$, using the last two valid molecules $(M_{i-1}^j,m_i^j)$ as input for the MolE module, METree-MPP predicts the property changes caused by the molecular pair. It learns the molecular evolutionary patterns, specifically the property changes caused by adding atoms and chemical bonds at different positions. 

\begin{table}[t!]
    \renewcommand\arraystretch{1.1} % 行间距
    \setlength{\tabcolsep}{2mm}{
    \resizebox{\columnwidth}{!}{%
    \begin{tabular}{ccccc}
    \hline
    \multirow{2}{*}{Dataset} & \multirow{2}{*}{Molecules} & \multicolumn{2}{c}{MEvoN} & \multirow{2}{*}{Max Path} \\ \cline{3-4}
     &  & Edges & Nodes &  \\ \hline
    QM7 \cite{QM7} & 6832 & 9095 & 6832 & 110 \\ \hline
    QM8 \cite{GDB17} & 21766 & 27068 & 21766 & 46 \\ \hline
    QM9 \cite{QM9} & 133885 & 165790 & 133330 & 253 \\ \hline
    \end{tabular}}}
    \caption{Overview of MEvoN construction on the QM7, QM8, and QM9 datasets.}
    \label{tab:qm_datasets_summary}
\end{table}

In the Evo branch, the molecular evolutionary pair's features $X_1$ and $X_2$ are extracted using the MolE. Then, the difference between these features $\Delta X =  X_2-X_1$ is computed. Subsequently, the difference feature $\Delta X$ is concatenated with the evolutionary path features $\mathbf{H}_{\text{EvoP}}$ and $\mathbf{H}_{\text{EvoL}}$ extracted by the EvoP  and EvoL modules. This can be expressed as:
\begin{align}
\left\{
\begin{aligned}
y_1^{\mathrm{pred}} &= \mathcal{F} (X_1), \\
y_2^{\mathrm{pred}} &= \mathcal{F} \left( \left[ \Delta X, X_1, X_2, \mathbf{H}_{\text{EvoP}}, \mathbf{H}_{\text{EvoL}} \right] \right),
\end{aligned}
\right.
\end{align}
\noindent where the multilayer perceptron is denoted as  $\mathcal{F}(\cdot)$. Finally, the loss function is defined as:
\begin{equation}
\mathcal{L} = \alpha \cdot \text{MSE}(y_1^{\text{pred}},  y^i_{l-1}) + \beta \cdot \text{MSE}(y_2^{\text{pred}}, y^i_{l-2} - y^i_{l-1}).
\end{equation}
\noindent where $\alpha$ and $\beta$ are hyperparameters for loss weights and the $\text{MSE}(\cdot)$ stands for mean squared error.

\begin{table*}[t!]
\renewcommand\arraystretch{1.15} % 行间距
\setlength{\tabcolsep}{5mm}{
\resizebox{\textwidth}{!}{%
\begin{tabular}{cccccccccc}
    \toprule
     & Property & $\varepsilon_{\mathrm{HOMO}}$ & $\varepsilon_{\mathrm{LUMO}}$ & $\Delta\varepsilon$ & $\mathrm{ZPVE}$ & $\mu$ & $\alpha $ & $\left \langle \mathit{R} \right \rangle ^2$ & $C_V$ \\ \cline{2-10}
    \multirow{-2}{*}{Methods} & Unit & eV & eV & eV & eV & D & bohr$^3$ & bohr$^2$ & cal/mol K \\ \hline
     & Original & 0.2539 & 0.1336 & 0.5928 & 0.3273 & 0.7943 & 2.2469 & 101.7469 & 1.5261 \\
     & Mol-branch & \textbf{0.1419} & \textbf{0.1207} & 0.2605 & 0.0649 & 0.6134 & 2.5860 & 68.2874 & 0.9806 \\
     & Evo-branch & 0.1453 & 0.1373 & \textbf{0.2579} & \textbf{0.0322} & \textbf{0.5825} & \textbf{0.7621} & \textbf{46.5656} & \textbf{0.3432} \\ 
    \multirow{-4}{*}{GCN} & (Improve.) & {\color[HTML]{CB0000} 44.11\%} & {\color[HTML]{CB0000} 9.64\%} & {\color[HTML]{CB0000} 56.50\%} & {\color[HTML]{CB0000} 90.16\%} & {\color[HTML]{CB0000} 26.67\%} & {\color[HTML]{CB0000} 66.08\%} & {\color[HTML]{CB0000} 54.23\%} & {\color[HTML]{CB0000} 77.51\%} \\ \hline
     & Original & 0.2174 & 0.1247 & 0.2916 & 0.1622 & 0.4981 & 2.0459 & 77.8414 & 1.0610 \\
     & Mol-branch & 0.1662 & \textbf{0.1207} & 0.2270 & 0.1458 & 0.5087 & 2.5720 & 56.3729 & 3.0892 \\
     & Evo-branch & \textbf{0.1662} & 0.1267 & \textbf{0.2265} & \textbf{0.0572} & \textbf{0.4893} & \textbf{0.8115} & \textbf{35.2868} & \textbf{0.5370} \\
    \multirow{-4}{*}{GIN} & (Improve.) & {\color[HTML]{CB0000} 23.57\%} & {\color[HTML]{CB0000} 3.19\%} & {\color[HTML]{CB0000} 22.32\%} & {\color[HTML]{CB0000} 64.75\%} & {\color[HTML]{CB0000} 1.77\%} & {\color[HTML]{CB0000} 60.34\%} & {\color[HTML]{CB0000} 54.67\%} & {\color[HTML]{CB0000} 49.39\%} \\ \hline
     & Original & 0.0772 & 0.0586 & 0.1066 & 0.0053 & 0.0972 & 0.1813 & 1.6577 & 0.0625 \\
     & Mol-branch & 0.0538 & \textbf{0.0565} & \textbf{0.0809} & \textbf{0.0044} & 0.0646 & 0.1368 & \textbf{1.3347} & \textbf{0.0547} \\
     & Evo-branch & \textbf{0.0534} & 0.0578 & 0.0820 & 0.0064 & \textbf{0.0644} & \textbf{0.1271} & 1.5003 & 0.0557 \\
     \multirow{-4}{*}{\makecell{SchNet \\ \cite{schnet}}} & (Improve.) & {\color[HTML]{CB0000} 30.73\%} & {\color[HTML]{CB0000} 3.61\%} & {\color[HTML]{CB0000} 24.06\%} & {\color[HTML]{CB0000} 17.60\%} & {\color[HTML]{CB0000} 33.74\%} & {\color[HTML]{CB0000} 29.92\%} & {\color[HTML]{CB0000} 19.48\%} & {\color[HTML]{CB0000} 10.88\%} \\ \hline
     & Original & 0.0924 & 0.0638 & 0.1232 & \textbf{0.0068} & 0.1034 & 0.2997 & 2.2417 & 0.1200 \\
     & Mol-branch & 0.0568 & 0.0533 & \textbf{0.0862} & 0.0105 & 0.0833 & 0.2268 & 2.1444 & 0.0918 \\
     & Evo-branch & \textbf{0.0559} & \textbf{0.0530} & 0.0862 & 0.0071 & \textbf{0.0825} & \textbf{0.2151} & \textbf{2.0802} & \textbf{0.0892} \\
    \multirow{-4}{*}{\makecell{ComENet \\ \cite{ComENet}}} & (Improve.) & {\color[HTML]{CB0000} 38.56\%} & {\color[HTML]{CB0000} 16.45\%} & {\color[HTML]{CB0000} 30.00\%} & {\color[HTML]{009901} -3.92\%} & {\color[HTML]{CB0000} 19.44\%} & {\color[HTML]{CB0000} 28.23\%} & {\color[HTML]{CB0000} 4.34\%} & {\color[HTML]{CB0000} 25.69\%} \\\bottomrule
    \end{tabular}}}
    \caption{Enhancement effect of MEvoN-MPP on four molecular representation models, evaluated using MAE.}
    \label{tab:comparison_qm9}
\end{table*}

\section{Experiments}

% \subsection{Evaluation Strategies and Metrics}

This study focuses on MEvoN-based molecular property prediction utilizing the QM7 \cite{QM7} and QM9 \cite{QM9} datasets, which provide extensive quantum chemical properties for molecular modeling and property prediction. The datasets were randomly split into training, validation, and test sets with a ratio of 8:1:1. For the QM7 experiments, seeds 38–42 were used, while for QM9, random seed 42 was employed. These regression tasks apply the mean absolute error (MAE) used as the performance metric. The default values of the loss weights $\alpha$ and $\beta$ are both 1.

% These properties encompass key aspects of molecular electronic structure, geometry, and thermodynamic properties, providing a comprehensive characterization of molecular behavior. 

% The entire experiment is conducted using the PyTorch framework, to predict these properties through the learning and modeling of molecular evolutionary trees, ultimately supporting molecular design and drug discovery processes.

% \begin{figure}[t!]
%     \centering
%     \includegraphics[width=0.999\linewidth]{evolution_graph_20_graph_0.3_v1.png}
%     \caption{Evolutionary graph representation based on structural similarity, where the graph distance is used to quantify the similarity between molecular nodes.}
%     \label{fig:enter-label}
% \end{figure}

% \begin{figure}[t!]
%     \centering
%     \includegraphics[width=0.999\linewidth]{evolution_graph_20_edit_distance_0.5_v1.png}
%     \caption{Evolutionary graph representation based on edit distance, illustrating the molecular transformations between nodes with a threshold of 0.5 for edit distance.}
%     \label{fig:enter-label}
% \end{figure}

\begin{figure}[t!]
    \centering
    \includegraphics[width=0.99\linewidth]{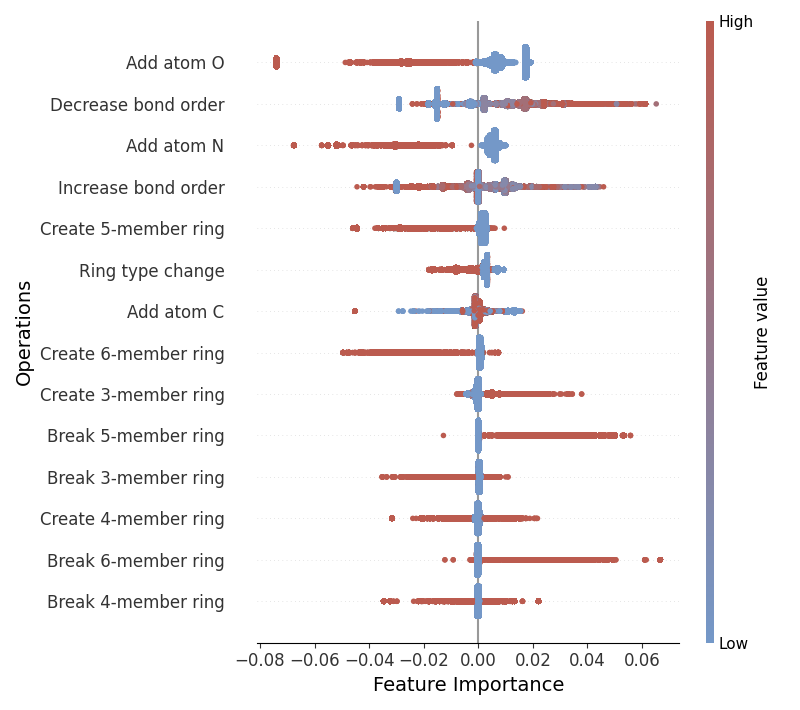}
    \caption{Importance of different mutation types in molecular evolution for the QM9 dataset with the GAP as the target property.  Each point represents a feature (i.e., mutation type) and its corresponding SHAP value. In this case, the color indicates the feature value and the position along the x-axis reflects the impact on the target property.}
    \label{fig:shap_summary}
\end{figure}

\subsection{Validating MEvoN}

To validate the MEvoN model's effectiveness, we constructed three evolutionary networks based on the QM7, QM8, and QM9 datasets. Figure \ref{tab:qm_datasets_summary} shows the number of molecules in the datasets, the number of edges and nodes in MEvoN, and the maximum number of paths per molecule. The networks were constructed using Algorithm \ref{alg:MEvoN} and the similarity thresholds were set to 0.3. Notably, 555 molecules from QM9 were excluded from the network because F-containing molecules (a total of 446) exhibited low similarity with most of the others, making it challenging to establish evolutionary relationships. Additionally, some N-containing cyclic structures showed low similarity to the main molecules and were also excluded. 

A detailed analysis of the QM9 dataset revealed that the mutations could be categorized into 14 types: adding C, N, and O atoms; increasing or decreasing bond order; forming or breaking 3/4/5/6-membered rings; and changing ring types. We observed the influence of different mutations on molecular relationships. Based on SHapley Additive exPlanations (SHAP) analysis \cite{SHAP}, we built a simplified regression model to quantify the impact of various mutations on molecular properties, using HOMO-LUMO gap (GAP) as an example. The results, shown in Figure \ref{fig:shap_summary}, reveal regular patterns in the evolutionary modifications’ effects on molecular properties. For instance, adding 5- and 6-membered rings, and introducing O and N atoms, generally decreases the GAP. Meanwhile, adding a 3-membered ring tends to increase the GAP. The effects of increasing or decreasing bond order are mutually exclusive, with an increase leading to GAP reduction. Furthermore, adding a single C atom has a minimal effect. These patterns are consistent with the theoretical quantum chemistry findings in various studies \cite{pope1999electronic}, providing strong evidence for the MEvoN method's capacity and significance in capturing molecular evolutionary relationships.

\subsection{Molecular Property Prediction Using MEvoN}

To validate our proposed MEvoN-MPP method's effectiveness across various molecular encoding architectures, we conducted property prediction experiments on the QM7 and QM9 datasets. For the QM9 experiments, we selected eight commonly used properties as the targets following the MolCLR method \cite{wang2022molecular}. The baselines included geometry-based method such as SchNet \cite{schnet} and ComENet \cite{ComENet}, along with traditional graph convolutional network (GCN) \cite{gcn} and graph isomorphism network (GIN) \cite{gin}. 

The results of the eight property prediction tasks on the QM9 dataset, shown in Table \ref{tab:comparison_qm9}, demonstrate that MEvoN effectively enhances molecular representations by an average of 32.3\%. The average performance improvements of GCN, GIN, SchNet, and ComENet are 53.11\%, 35.00\%, 21.25\%, and 19.85\%, respectively. The QM7 results, shown in Table \ref{tab:comparison_qm7}, indicate that our GCN-based model competes with leading methods like D-MPNN \cite{DMPNN}, Attentive FP \cite{AttFP}, GROVER \cite{GROVER}, GEM \cite{GEM}, PretrainGNN \cite{PretrainGNN}, and Uni-Mol \cite{Uni-Mol}. Notably, these methods rely on large-scale pretraining followed by fine-tuning on the QM7 dataset. For example, Uni-Mol is pretrained on a database containing 19 million molecules and 209 million conformations, whereas our method is trained and evaluated only on the QM7 dataset (with less than 7,000 molecules). Comparing the results across different methods reveals that our method significantly improves molecular property prediction accuracy, surpassing traditional and pretraining-based models.

\subsection{Ablation Experiments}

% 为探究MEvoN-MPP各模块的作用，我们设计了消融实验，以回答以下问题：EvoP模块和EvoL是否能够有效增强MEvoN对分子表征的能力？它们各自的贡献有多大？

% 实验中选择QM9数据集的GAP属性作为预测目标，消融实验结果如表格X所示。仅使用MolE进行预测时，MAE为0.291。引入EvoP模块或EvoL单独运行时，MAE分别降低了约15%。当EvoP模块和EvoL共同使用时，与仅使用MolE相比，性能提升了约30%，显著表明这两个模块对模型性能的增强作用。

\begin{table}[t!]
\centering
\renewcommand\arraystretch{1.1} % 行间距
\setlength{\tabcolsep}{5mm}{
\resizebox{\columnwidth}{!}{%
\begin{tabular}{cccc}
\toprule
Methods & Result(MAE) & Methods & Result(MAE) \\
\midrule
D-MPNN & 103.5(8.6) & PretrainGNN & 113.2(0.6) \\
GROVER$_{\text{base}}$ & 94.5(3.8) &  MolCLR & 66.8(2.3) \\
GROVER$_{\text{large}}$ & 92.0(0.9) &  GEM & 58.9(0.8) \\
Attentive FP & 72.0(2.7) & Uni-Mol & \textbf{41.8(0.2)} \\ \hline
MEvoN-MPP$_{\text{gcn}}$ & \underline{45.9(3.4)} & MEvoN-MPP$_{\text{gin}}$ & 65.6(6.3) \\
\toprule
\end{tabular}}}
\caption{Comparison of MAE results for different models on QM7 dataset.}
\label{tab:comparison_qm7}
\end{table}

\begin{table}[t!]
    \centering
    \renewcommand\arraystretch{1.05} % 行间距
    \setlength{\tabcolsep}{6mm}{
    \resizebox{\columnwidth}{!}{%
    \begin{tabular}{ccccc}
        \toprule
        \multirow{2}{*}{EvoP} & \multirow{2}{*}{EvoL}  & \multirow{2}{*}{MolE} & \multicolumn{2}{c}{Result(MAE)} \\
        \cline{4-5}
        & & & Mol-branch & Evo-branch \\
        \midrule
        \ding{55} & \ding{55} & \ding{51} & 0.2916 & - \\
        \ding{51} & \ding{55} & \ding{51} & 0.9809 & 0.5187 \\
        \ding{55} & \ding{51} & \ding{51} & 0.4051 & 0.3521 \\
        \ding{51} & \ding{51} & \ding{51} & \textbf{0.2270} & \textbf{0.2265} \\
        \bottomrule
    \end{tabular}
    }
    }
    \caption{Ablation study of MEvoN-MPP on the QM9 dataset for the GAP property.}
    \label{tab:ablation}
\end{table}

% To investigate the contributions of each module in MEvoN-MPP, we designed an ablation study to address the following questions: 
% Do the EvoP module and EvoL effectively enhance MEvoN's ability to represent molecular features? What are their respective impacts?

% In this study, the GAP property from the QM9 dataset was selected as the target for prediction, and the results of the ablation experiments are presented in Table \ref{tab:ablation}. When only MolE was used for prediction, the MAE was 0.291. Introducing either the EvoP module or EvoL individually reduced the MAE by approximately 15\%. When both the EvoP module and EvoL were employed together, the performance improved by about 30\% compared to using only MolE, clearly demonstrating the significant enhancement provided by these two modules.

% In this study, we selected the GAP property from the QM9 dataset as the prediction target. 

To investigate the contributions of different modules in MEvoN-MPP, we conducted an ablation study to evaluate the EvoP and EvoL modules' ability to leverage MEvoN's molecular feature representations and impact exploration capacity effectively. The ablation study's results are presented in Table \ref{tab:ablation}. When only MolE was used for prediction, the MAE was 0.2916. Introducing the EvoP or EvoL modules individually increased the MAE to approximately 0.98 and 0.40, respectively. This indicates that these modules are not effective when applied independently. However, when the EvoP and EvoL modules applied together, model performance improved by approximately 22.3\% compared to using only MolE. This demonstrates that the collaboration between the EvoP and EvoL modules significantly enhances the model's ability to predict molecular properties. Furthermore, the two modules effectively integrate the molecule's local and global features by combining path and label sequence encoding. This fusion mechanism enables the model to focus on structural changes at key positions while capturing the molecule's overall evolutionary trends, leading to a more comprehensive and precise representation of molecular features.

\subsection{Hyperparameter Experiments}

% The table shows the results of ablation experiments for predicting Homo (HOMO energy) using models with varying path numbers (Path Num). The evaluation metrics include R (Coefficient of Determination), PCC (Pearson Correlation Coefficient), MSE (Mean Squared Error), MAE (Mean Absolute Error), and the number of epochs (Epoch). These results demonstrate how model performance changes with the number of paths.

\begin{figure}[t!]
    \centering
    \includegraphics[width=0.95\linewidth]{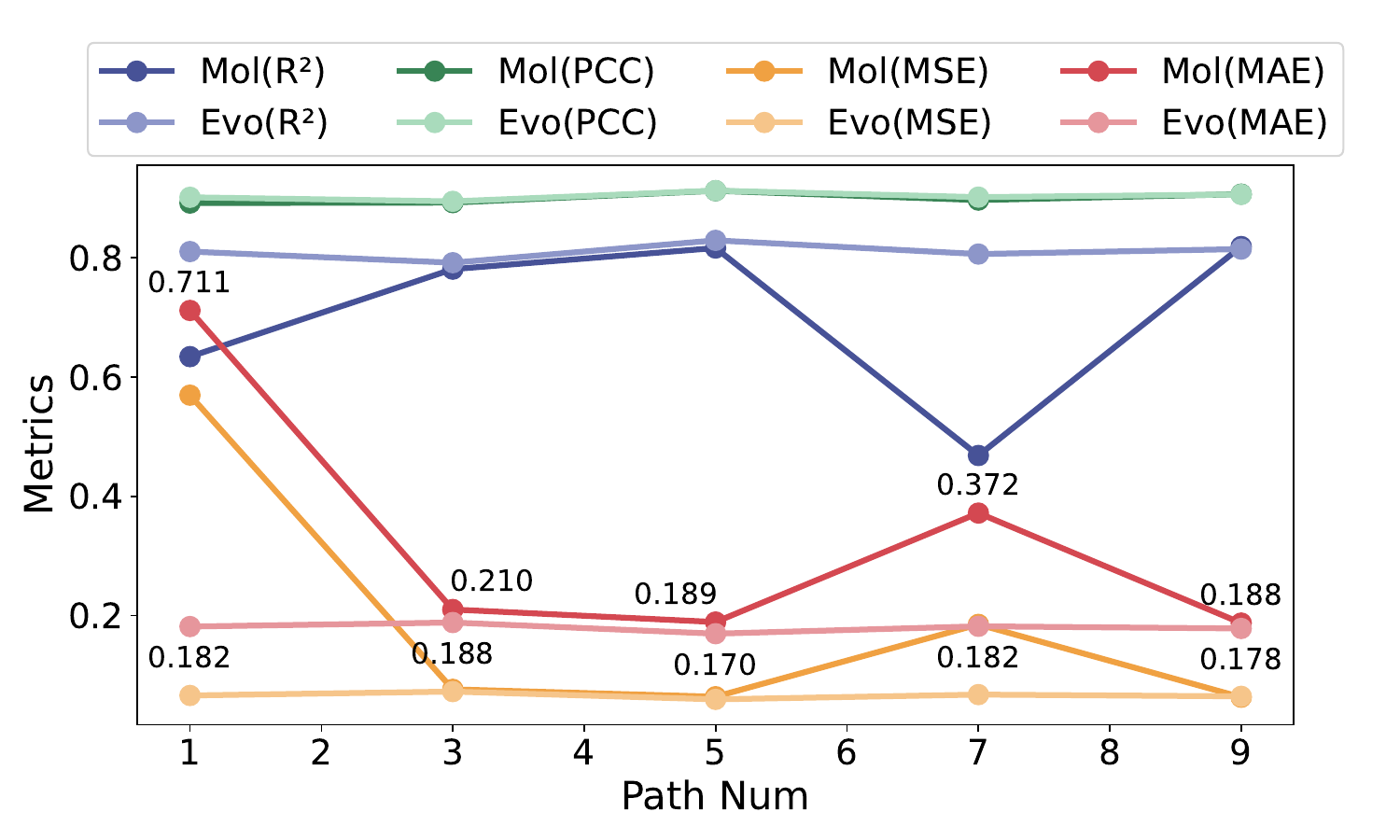}
    \caption{Hyperparameter experiment exploring the impact of different path numbers on evolutionary path representation. The evaluation metrics included the determination coefficient(R$^2$), Pearson correlation coefficient (PCC), mean squared error (MSE), and MAE.}
    \label{fig:Hyperparameter}
\end{figure}

In our MEvoN model, each molecule typically has more than one evolutionary path. To investigate the impact of embedding different evolutionary path numbers on molecular representation, we conducted hyperparameter experiments. As shown in Figure \ref{fig:Hyperparameter}, we evaluated the $\varepsilon_{\mathrm{HOMO}}$ property on the QM9 dataset with path numbers K set to 1, 3, 5, 7, and 9, using four regression metrics. To ensure fair comparisons, the training epoch was fixed at 150 across all experiments. The results indicate that K has a significant effect on molecular representation. Specifically, when K is set between 3 and 5, the model demonstrates stable performance, with few errors and high prediction accuracy during regression tasks. Conversely, insufficient or excessive paths lead to instability. Obtaining insufficient paths may result in a failure to capture the diversity and complexity of molecular evolution, reducing prediction accuracy. In contrast, an excessive number of paths increases computational complexity, slowing model convergence and resulting in decreased performance during comparative evaluations. 

% Each molecule is assigned several evolutionary sequence of length $l$, which represents its evolutionary history. The complexity of each group is $\left(\frac{N}{\text{GroupNum}}\right)^2$, where $\text{GroupNum}$ is the number of groups. Given that we have $W$ groups, the overall algorithm complexity is $\mathcal{O}\left(W \cdot \left(\frac{N}{\text{GroupNum}}\right)^2\right)$.

% The number of evolutionary paths reflects both the diversity of molecular evolution directions and the varied impacts of mutations on molecular properties. Each path represents the evolutionary process under different mutation operations, enabling the model to better capture the intricate relationships between molecular features and properties. A balanced number of paths effectively enhances the model’s ability to capture structural variations while maintaining prediction stability.

\section{Conclusion}

This paper introduces a novel molecular representation method based on the MEvoN. By simulating the evolutionary pathway from ancestral to current structures, MEvoN captures dynamic, multi-level features reflecting molecular structural changes. When combined with traditional encoding methods, MEvoN enhances molecular representation for downstream tasks. To validate its effectiveness, we applied MEvoN to molecular property prediction tasks, experimenting on eight sub-tasks from the QM7 and QM9 datasets and using four encoding methods. The results demonstrate a 32.3\% average performance improvement. Therefore, the MEvoN effectively captures structural variations, deepening our understanding of the relationship between molecular evolution and properties, with promising applications in drug discovery and materials optimization.

% The results show a 16% average performance improvement. MEvoN captures structural variations, deepening our understanding of molecular evolution and its applications in drug discovery and materials optimization.

\appendix

\clearpage
% \section*{Ethical Statement}

% There are no ethical issues.

% \section*{Acknowledgments}

% The preparation of these instructions and the \LaTeX{} and Bib\TeX{}
% files that implement them was supported by Schlumberger Palo Alto
% Research, AT\&T Bell Laboratories, and Morgan Kaufmann Publishers.
% Preparation of the Microsoft Word file was supported by IJCAI.  An
% early version of this document was created by Shirley Jowell and Peter
% F. Patel-Schneider.  It was subsequently modified by Jennifer
% Ballentine, Thomas Dean, Bernhard Nebel, Daniel Pagenstecher,
% Kurt Steinkraus, Toby Walsh, Carles Sierra, Marc Pujol-Gonzalez,
% Francisco Cruz-Mencia and Edith Elkind.

\bibliographystyle{named}
\bibliography{ijcai25}

\end{document}